\documentclass[aps,pra,twocolumn,groupedaddress]{revtex4-1}
\usepackage{physics}
\usepackage{graphicx} 
\newcommand{\ii}{{\mathrm{i}}}
\newcommand{\ee}{{\mathrm{e}}}
\begin{document}
\title{Degenerate spontaneous parametric down-conversion in nonlinear metasurfaces}
\author{Tetsuyuki Ochiai}
\affiliation{Research Center for Electronic and Optical Materials, National Institute for Materials Science (NIMS), Tsukuba 305-0044, Japan}
\date{\today}

\begin{abstract}
We propose a simple scheme of degenerate spontaneous parametric down-conversion (SPDC) in nonlinear metasurfaces or photonic crystal slabs with quasi-guided modes.   
It employs a band crossing between P- and S-polarized quasi-guided mode bands inside the light cone and a selection rule in the conversion efficiency of the SPDC.   
The efficiency can be evaluated fully classically via the inverse process of noncollinear second-harmonic generation (SHG). As a toy model, we study the SPDC and SHG in a monolayer of noncentrosymmetric spheres and confirm that the scenario works well to enhance the SPDC.  
\end{abstract}

\maketitle

\section{Introduction}
The spontaneous parametric down-conversion (SPDC) is a quantum phenomenon of generating two photons with angular frequency $\omega_1$ and $\omega_2$ by a pump light of angular frequency $\omega_1+\omega_2$ \cite{PhysRevA.31.2409}. 
It becomes a source of entangled photon pairs \cite{PhysRevA.50.5122}, so that the SPDC can be used in quantum information, such as the Bell inequality test experiment \cite{PhysRevLett.61.2921,PhysRevLett.81.3563}.

Compared with other entangled photon sources, such as the cascade emission from three-level atoms or biexcitons in quantum dots  \cite{PhysRevLett.18.575,PhysRevLett.99.266802,PhysRevB.88.041306}, the SPDC has its advantages and disadvantages.  
One advantage is it's relatively easy to set up. The disadvantages include (in principle) low generation efficiency, lack of determinism, and need for filtering. 
These properties are tied with the SPDC being a nonlinear optical process.

As a second-order nonlinear optical process, the SPDC is observed typically in uniaxial and noncentrosymmetric media \cite{boyd-nonlinear-book}. There, the phase matching via the birefringe is crucial.  
For instance, the phase matching of type II gives the polarization entanglement between the ordinary and extraordinary waves with orthogonal polarizations.

Optical media that are uniaxial and noncentrosymmetric are limited.  Relaxing these two properties are essential for practical applications involving the second-order optical nonlinearity.  One crucial direction here is to employ the quasi-phase matching by introducing an artificial spatial periodicity in the system \cite{fejer1992quasi,myers1995quasi,PhysRevLett.93.040504,PhysRevLett.100.183601}. 
There, we can be free from the birefringe of the bulk materials.   
Another important direction is to employ surface modes in, even centrosymmetric media \cite{PhysRevB.27.1965,valev2012characterization}. Since any material breaks the centrosymmetry at the surface, the system can have $\chi^{(2)}$ there. 
Combining possible surface modes with surface (or bulk) $\chi^{(2)}$, we can expect a significant enhancement of the second-order optical nonlinearity even in optically thin specimens.

Recently, metasurfaces or photonic crystal slabs including nonlinear materials have attracted growing interest as a platform for enhanced nonlinear optical processes \cite{minovich2015functional}. This trend is  in the latter direction.   
In particular, the so-called bound states in the continuum (BICs) \cite{OHTAKA:I::40:p425-428:1981,PhysRevB.58.6920,Paddon:Y::61:p2090-2101:2000,Ochiai:S::63:p125107:2001,hsu2013observation} or, more generally, quasi-guided modes  \cite{Fan:J::65:p235112:2002,Tikhodeev:Y:M:G:I::66:p045102:2002} are vital items there. 
These modes have high (ideally infinite) quality factors so that the light-matter interaction involving them becomes very strong \cite{kang2023applications}.      
So far, various functions such as lasing \cite{kodigala2017lasing,ha2018directional,hwang2021ultralow}, second-harmonic generation (SHG) \cite{ochiai2017enhanced,bernhardt2020quasi,PhysRevLett.123.253901,PhysRevB.106.125411}, and Kerr effect \cite{PhysRevB.97.224309,PhysRevA.102.033511,kang2023nonlinear}, have been demonstrated via nonlinear metasurfaces. 
However, the study of SPDC in nonlinear metasurfaces is still limited \cite{Wang:19,jin2021efficient,parry2021enhanced,santiago2022resonant,mazzanti2022enhanced}. 

For instance, previous studies on the SPDC in two-dimensional (2D) metasurfaces rely on deformations (in geometry) of symmetry-protected BICs at the $\Gamma$ point so that the emission angles of the signal and idler photons are close to the normal direction of the metasurfaces \cite{parry2021enhanced,santiago2022resonant}. There is a broad tunability of relevant quality factors in such cases. 
The tunability is reduced in one-dimensional (1D) metasurfaces  \cite{jin2021efficient,mazzanti2022enhanced}.

Instead of such a deformation, we here consider plain 2D structures and a conventional phenomenon of band crossings at a generic $\vb*{k}$ point on a mirror axis. 
This setting allows us to imitate the conventional SPDC scheme of the type II phase matching even in isotropic but noncentrosymmetric media.

In this paper, we present a theoretical analysis of the SPDC, taking account of metasurface structures and their spatial symmetries in a first-principles manner. 
We present a simple scenario of the degenerate SPDC and polarization entanglement in metasurfaces with a certain photonic band structure. 
The essential quantity is the conversion efficiency from the pump light to biphoton states. The spatial symmetries of the metasurface and the symmetry of $\chi^{(2)}$ yields a selection rule in the conversion efficiency factor.     
It restricts possible combinations of the biphoton states and pump-light polarization.  
Moreover, the efficiency factor can be strongly enhanced by the excitation of quasi-guided modes in the metasurface.  
We demonstrate these features in terms of a classical electromagnetic calculation of the reverse process, namely, the sum-frequency generation (SFG) \cite{lambert2005implementing,PhysRevB.54.5732,PhysRevB.54.5742} or the noncollinear SHG.

This paper is organized as follows. In Sec. II, we summarize several formulas of the SPDC in terms of an eigenmode expansion of the quantized radiation field. Section III is devoted to presenting the reverse process, namely, the SFG, in a fully classical approach. 
In Sec. IV, we present symmetry properties of the conversion efficiency factor in the SPDC and resulting polarization entanglement in nonlinear metasurfaces 
assuming a mirror symmetry.   
In Sec. V, we present a simulation of the conversion efficiency in a monolayer of noncentrosymmetric spheres as a toy model.  
Finally, in Sec. VI, summary and discussion are given.

\section{spontaneous parametric down conversion}
We first consider a generic photonic system with noncentrosymmetric media. 
The Hamiltonian ${\cal H}$ of the radiation field including the second-order optical nonlinearity becomes \cite{drummond2014quantum}
\begin{align}
&{\cal H} = {\cal H}_0+{\cal H}_1,\\ &{\cal H}_0=\int \dd^3x\qty( \frac{\mu_0}{2}\vb*{H}^2+\frac{1}{2\epsilon_0}\vb*{D}\tensor{\eta}^{(1)}\vb*{D}), \\
&{\cal H}_1=\int \dd^3x \frac{1}{3\epsilon_0^2}\eta_{ijk}^{(2)}D_iD_jD_k,\\
&\tensor{\eta}^{(1)}= (\tensor{\epsilon}^{(1)})^{-1}, \quad \tensor{\epsilon}^{(1)}=\tensor{1}+\tensor{\chi}^{(1)}, \\
&\eta_{ijk}^{(2)}=-\chi_{lmn}^{(2)}\eta_{il}^{(1)}	\eta_{mj}^{(1)}	\eta_{nk}^{(1)}, 	\label{Eq:eta2}
\end{align}
where $\vb*{D}$ and $\vb*{H}$ are the electric displacement and magnetic fields, respectively, $\mu_0$ and $\epsilon_0$ are the vacuum permeability and permittivity, respectively, and   
$\chi^{(1)}$ and $\chi^{(2)}$ are space-dependent linear and second-order electric susceptibilities, respectively, of the media. That is, 
\begin{align}
P_i = \epsilon_0(\chi_{ij}^{(1)}E_j + \chi_{ijk}^{(2)}E_jE_k),\label{Eq:polarization}
\end{align}	
being $\vb*{P}$ and $\vb*{E}$ the electric polarization and electric field, respectively, with constitutive relation $\vb*{D}=\epsilon_0\vb*{E}+\vb*{P}$.

We introduce the dual vector potential $\vb*{\Lambda}$ with the "Coulomb" gauge ($\div\vb*{\Lambda}=0$) as  
\begin{align}
	\vb*{D}=\curl\vb*{\Lambda}, \quad \vb*{H}=\pdv{\vb*{\Lambda}}{t}.	
\end{align}
Using the eigenmodes of $\vb*{\Lambda}$ of the linear Maxwell equation, namely, 
\begin{align}
&\nabla\times \qty(\tensor{\eta}^{(1)}(\vb*{x})\nabla\times \vb*{\Lambda}_\alpha(\vb*{x}))=\frac{\omega_\alpha^2}{c^2}	\vb*{\Lambda}_\alpha(\vb*{x}),\\
&\int \dd^3x \vb*{\Lambda}_\alpha^*(\vb*{x})\cdot\vb*{\Lambda}_\beta(\vb*{x})=\delta_{\alpha\beta},
\end{align}
the radiation field can be expanded as 
\begin{align}
&\vb*{D}(\vb*{x},t)=\sum_\alpha \sqrt{\frac{\hbar}{2\mu_0\omega_\alpha}}(
a_\alpha(t)\curl \vb*{\Lambda}_\alpha(\vb*{x})\nonumber\\
&\hskip85pt +a_\alpha^\dagger(t)\curl \vb*{\Lambda}_\alpha^*(\vb*{x})
	),\label{Eq:Dquantum}\\
&\vb*{H}(\vb*{x},t)=-\ii\sum_\alpha \sqrt{\frac{\hbar\omega_\alpha}{2\mu_0}}\qty(
a_\alpha(t)\vb*{\Lambda}_\alpha(\vb*{x})-
a_\alpha^\dagger(t)\vb*{\Lambda}_\alpha^*(\vb*{x})
), 
\end{align}
where index $\alpha$ represents an eigenstate, $\omega_\alpha$ is its eigenfrequency, and $\vb*{\Lambda}_\alpha$ is its normalized eigenfunction. 
The unperturbed Hamiltonian ${\cal H}_0$ then 
becomes 
\begin{align}
&{\cal H}_0=\sum_\alpha \frac{\hbar\omega_\alpha}{2}\qty (a_\alpha^\dagger(t) a_\alpha(t) + a_\alpha(t) 
a_\alpha^\dagger(t) 
),\\
&[ a_\alpha(t),a_\beta^\dagger(t) ]=\delta_{\alpha\beta},  	
\end{align}
assuming the time-reversal symmetry: 
\begin{align}
\vb*{\Lambda}_\alpha^*=\ee^{\ii\lambda_\alpha}\vb*{\Lambda}_{-\alpha},\quad \omega_\alpha=\omega_{-\alpha}.
\end{align}	
In the interaction picture, we have 
\begin{align}
a_\alpha(t)=a_\alpha \ee^{-\ii\omega_\alpha t}.	
\end{align}
The state vector is time-developed as
\begin{align} 
|\psi(t)\rangle = {\cal T}\ee^{-\frac{\ii}{\hbar}\int_{-\infty}^t \dd t' {\cal H}_1(t') }	|\psi(-\infty)\rangle 
\end{align} 
where $\cal{T}$ represents the time-ordering product.

Let us consider a strong continuous wave (CW) pump light of an eigenstate $\alpha=p$ is incident on the system. Its fluctuation can be safely neglected so that the interaction Hamiltonian ${\cal H}_1$ relevant to the SPDC is 
\begin{align}
&{\cal H}_1\to \int \dd^3x \frac{1}{\epsilon_0^2}\eta_{ijk}^{(2)} D_i^pD_j^fD_k^f, \\
&\vb*{D}^p(\vb*{x},t)= \sqrt{\frac{\hbar}{2\mu_0\omega_p}}\qty(
\beta_p \ee^{-\ii\omega_p t}\curl \vb*{\Lambda}_p(\vb*{x})+\mathrm{c.c.}
),
\end{align}
where $\vb*{D}^{p}$ is the classical-pump field, $\vb*{D}^{f}$ is the quantum-fluctuation expressed by Eq. \eqref{Eq:Dquantum}, and c.c. represents the complex conjugate.  
Here, we assume the permutation symmetry concerning the indices of $\eta_{ijk}^{(2)}$ holds, assuming a nondispersive $\chi^{(2)}$.       

In the first-order perturbation, if we start from the vacuum state $|0\rangle$ at time $t=-\infty$, we then obtain the biphoton states after a long time: 
\begin{align}
&|\psi(\infty)\rangle \simeq |0\rangle 
+\frac{1}{\ii\hbar}\qty(\frac{\hbar}{2\mu_0})^{\frac{3}{2}}\beta_p\sum_{\alpha_1,\alpha_2}|1_{\alpha_1}\otimes 1_{\alpha_2}\rangle \nonumber \\ 
&\hskip50pt\times \frac{F_{p\alpha_1\alpha_2}}{\sqrt{\omega_p\omega_{\alpha_1}\omega_{\alpha_2}}}
2\pi\delta(\omega_p-\omega_{\alpha_1}-\omega_{\alpha_2}),\\
&F_{p\alpha_1\alpha_2}=\int \dd^3x \eta_{ijk}^{(2)}
(\curl\vb*{\Lambda}_p)_i(\curl\vb*{\Lambda}_{\alpha_1}^*)_j\nonumber\\
&\hskip100pt \times (\curl\vb*{\Lambda}_{\alpha_2}^*)_k, \label{Eq:conveff}
\end{align}	
where $|1_\alpha\rangle$ represents the one photon state with eigenstate $\alpha$. 
The second term is thus a superposition of entangled biphoton states between eigenstates $\alpha_1$ and $\alpha_2$. 
The conversion efficiency from the pump light to the biphoton state is represented by 
factor $F_{p\alpha_1\alpha_2}$.

\section{Sum-frequency generation}

To evaluate the SPDC, we need the factor 
$F_{p\alpha_1\alpha_2}$. The direct calculation of it according to Eq. \eqref{Eq:conveff} requires a detailed analysis of the eigenmode profiles. Instead, we consider the reverse process of the SPDC, namely, the SFG, in which the same factor emerges as we will see.   
Since the SFG does not need a quantum approach, we can access the factor fully classically in a perturbation scheme.

Suppose that a incident light of eigenstates $\alpha_1$ and $\alpha_2$ is impinging to the system:
\begin{align}
	&\vb*{D}^{(1)}(\vb*{x},t)=\Re\left[
	\sqrt{\frac{2\hbar}{\mu_0\omega_{\alpha_1}}}\beta_1 \curl\vb*{\Lambda}_{\alpha_1}(\vb*{x})\ee^{-\ii\omega_{\alpha_1}t} \right. \nonumber\\ 
&\hskip65pt \left.	+\sqrt{\frac{2\hbar}{\mu_0\omega_{\alpha_2}}}\beta_2  
  \curl\vb*{\Lambda}_{\alpha_2}(\vb*{x})\ee^{-\ii\omega_{\alpha_2}t} 
	\right],\\
	&\vb*{H}^{(1)}(\vb*{x},t)=\Re\left[
	-\ii\sqrt{\frac{2\hbar\omega_{\alpha_1}}{\mu_0}}\beta_1
	\vb*{\Lambda}_{\alpha_1}(\vb*{x})\ee^{-\ii\omega_{\alpha_1}t}\right. \nonumber\\
&\hskip70pt \left.
	 -\ii\sqrt{\frac{2\hbar\omega_{\alpha_2}}{\mu_0}}\beta_2 \vb*{\Lambda}_{\alpha_2}(\vb*{x})\ee^{-\ii\omega_{\alpha_2}t} 
	\right],		
\end{align}
In the perturbative viewpoint, this field forms the nonlinear polarization $\vb*{P}^{(\mathrm{NL})}$ through Eq. \eqref{Eq:polarization} as 
\begin{align}
&P_i^{(\mathrm{NL})}(\vb*{x},t)=\epsilon_0\chi_{ijk}^{(2)} (\vb*{x})E_j^{(1)}(\vb*{x},t)	E_k^{(1)}(\vb*{x},t),\\
&\vb*{E}^{(1)}(\vb*{x},t)=\frac{1}{\epsilon_0}\tensor{\eta}^{(1)}(\vb*{x})
	\vb*{D}^{(1)}(\vb*{x},t). 	
\end{align}
The nonlinear polarization has several frequency components. 
The component of angular frequency $\omega=\omega_{\alpha_1}+\omega_{\alpha_2}$ is given by 
\begin{align}
	&\vb*{P}^{(\mathrm{NL})}(\vb*{x},t)\to \Re[\vb*{P}_\omega^{(\mathrm{NL})}(\vb*{x})\ee^{-\ii\omega t}], \\
	&(\vb*{P}_\omega^{(\mathrm{NL})}(\vb*{x}))_i=-2\hbar c^2\frac{\beta_1\beta_2}{\sqrt{\omega_{\alpha_1}\omega_{\alpha_2}}}\epsilon_{ij}^{(1)}(\vb*{x}) \eta_{jkl}^{(2)}(\vb*{x})\nonumber\\
	&\hskip50pt \times (\curl\vb*{\Lambda}_{\alpha_1}(\vb*{x}))_k
	(\curl\vb*{\Lambda}_{\alpha_2}(\vb*{x}))_l. 
\end{align}
This polarization becomes the source of the SFG.

The dual vector potential induced by the source satisfies 
\begin{align}
	\curl(\tensor{\eta}^{(1)}\curl \vb*{\Lambda}_\omega^{(2)})-\frac{\omega^2}{c^2}\vb*{\Lambda}_\omega^{(2)}
	=\curl	(\tensor{\eta}^{(1)}\vb*{P}_\omega^{(\mathrm{NL})}).
\end{align} 
By using the eigenmode expansion, the equation is solved as 
\begin{align}
	&\vb*{\Lambda}_\omega^{(2)}(\vb*{x})=\sum_\alpha c_\alpha^{(2)} \vb*{\Lambda}_\alpha(\vb*{x}), \\	
	&c_\alpha^{(2)}=2\hbar c^4\beta_1\beta_2\frac{F_{\alpha\alpha_1\alpha_2}^*}{(\omega^2-\omega_\alpha^2)\sqrt{\omega_{\alpha_1}\omega_{\alpha_2}}}. 
\end{align} 
The radiation flux of the SFG is given by 
\begin{align}
	&\int \dd\vb*{S}\cdot\frac{1}{2}\Re[\vb*{E}_\omega^{(2)*}\times \vb*{H}_\omega^{(2)}]\nonumber\\
	&\hskip10pt =\frac{\pi\hbar^2c^6}{\epsilon_0}|\beta_1\beta_2|^2\sum_\alpha \delta(\omega_\alpha-\omega_{\alpha_1}-\omega_{\alpha_2})\frac{|F_{\alpha\alpha_1\alpha_2}|^2}{\omega_{\alpha_1}\omega_{\alpha_2}}. \label{Eq:flux}
\end{align}
If the frequency-matched eigenstate is unique, we can access factor $|F_{\alpha\alpha_1\alpha_2}|^2$ by fully classically calculating the radiation flux.

\section{Selection rule in metasurfaces}

In what follows, we consider a nonlinear metasurface that is periodic in the $xy$ plane and is finite in the $z$ direction.
We do not care about the details of the system, but we impose that the system has a mirror symmetry on a particular axis of the system and the time-reversal symmetry. We further assume the mirror axis bisects the unit cell (UC).

According to the Bloch theorem, the system is characterized by a 2D Bloch momentum $\vb*{k}$. 
For $\vb*{k}$ on the mirror axis, say the $x$ axis, the eigenmodes are classified according to the parity (of the $y$ inversion) concerning the axis. 
Then,  we have 
\begin{align}
&-\hat{\sigma}_y \vb*{\Lambda}_{\tilde{\alpha}\vb*{k}}(\hat{\sigma}_y\vb*{x}) =\pm\vb*{\Lambda}_{\tilde{\alpha}\vb*{k}}(\vb*{x}),\\
&\vb*{\Lambda}_{\tilde{\alpha}\vb*{k}}(\vb*{x})=\ee^{\ii\vb*{k}\cdot\vb*{x}_\|} \vb*{u}_{\tilde{\alpha}\vb*{k}}(\vb*{x}),\\
&\vb*{u}_{\tilde{\alpha}\vb*{k}}(\vb*{x}+\vb*{X})=\vb*{u}_{\tilde{\alpha}\vb*{k}}(\vb*{x}),
\end{align}
where $\hat{\sigma}_y$ is the $y$ inversion operator,   
$\alpha=(\tilde{\alpha}\vb*{k})$, and $\vb*{X}$ is a 2D real lattice vector.  
Thus, we have 
\begin{align}
&F_{p\alpha_1\alpha_2}=\sum_{\vb*{X}}\ee^{\ii(\vb*{k}_p-\vb*{k}_{\alpha_1}-\vb*{k}_{\alpha_2})\cdot{\vb*{X}}} \tilde{F}_{p\alpha_1\alpha_2},\\
&\tilde{F}_{p\alpha_1\alpha_2}=\int_\mathrm{UC} \dd^3x \eta_{ijk}^{(2)}
(\curl\vb*{u}_{p\vb*{k}_p})_i(\curl\vb*{u}_{\tilde{\alpha}_1\vb*{k}_{\alpha_1}}^*)_j\nonumber\\
&\hskip100pt \times (\curl\vb*{u}_{\tilde{\alpha}_2\vb*{k}_{\alpha_2}}^*)_k,
\end{align}
The 2D phase matching requires $\vb*{k}_p=\vb*{k}_{\alpha_1}+\vb*{k}_{\alpha_2}$.

Suppose that the pump light is incident to the metasurface in the normal direction ($\vb*{k}_p=0$). It is then possible to produce the biphoton state having respective Bloch momenta of $\vb*{k}$  and $-\vb*{k}$ both on the mirror axis.
 In this case, the pump, signal, and idler states are classified according to the parity on the mirror axis. This property constrains the biphoton states regarding the parity. 
Moreover, the parity is directly related to the polarization of the biphoton state in the far field. 
Namely, if the signal (idler) state has an even parity, then it is P-polarized in the far field. The odd parity state becomes S-polarized in the far field, provided no open diffraction channel exists.

\begin{figure}
	\includegraphics[width=0.45\textwidth]{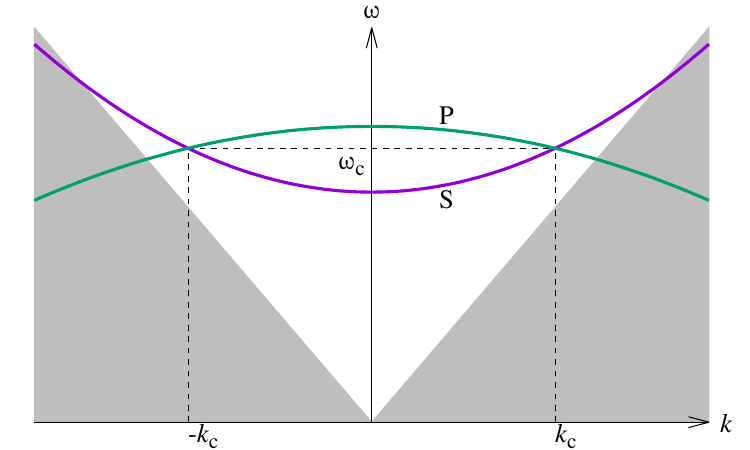}
	\caption{\label{Fig:schematic} 
		Schematic band structure for the SPDC. On a mirror axis of the system, the photonic band structure is classified into P- and S-polarized bands in the far field. At the crossing points, the selection rule in the conversion-efficiency factor of the SPDC results in a polarization entanglement between signal and idler photons for a pump light of the normal incidence. The signal and idler photons are degenerate with angular frequency $\omega_c$, and the momentum $\pm \vb*{k}_c$.  The gray area is the region outside the light cone.  
	}
\end{figure}

We here consider the case that two quasi-guided photonic bands of respective even and odd parities intersect at $\pm \vb*{k}_c$ as shown in Fig. \ref{Fig:schematic}. 
The eigenfrequency there is denoted as $\omega_c$. We assume  the pump frequency $\omega_p$ is twice the eigenfrequency, $\omega_p=2\omega_c$. 
Then, both the even and odd parity eigenstates of $\omega_c$ can be generated.

If the pump light is $x$-polarized and off-resonant, the nonvanishing 
$\eta^{(2)}_{ijk}$ components that contribute to  $F_{p\alpha_1\alpha_2}$ is 
\begin{align}
&(ijk)=(xxx), (xxz), (xzx), (xzz), (xyy)
\end{align}
for (signal,idler) being (P,P) or (S,S) polarized, and 
\begin{align}
&(ijk)=(xxy),(xyx),(xyz),(xzy) 
\end{align}
for (signal,idler) being (P,S) or (S,P) polarized.  
If the pump light is $y$-polarized, the nonvanishing 
$\eta^{(2)}_{ijk}$ components are 
\begin{align}
	&(ijk)=(yxx), (yxz), (yzx), (yzz), (yyy)
\end{align}
for (signal,idler)=(P,P) or (S,S), and 
\begin{align}
	&(ijk)=(yxy),(yyx),(yyz),(yzy) 
\end{align}
for (signal,idler)=(P,S) or (S,P).

In a simple noncentrosymmetric material whose crystal symmetry is $T_d$, the allowed nonzero component of $\chi^{(2)}$ is 
\begin{align}
(ijk)=(xyz),(xzy),(yzx),(yxz),(zxy),(zyx).
\end{align} 
 All the components are equal in $\chi^{(2)}$ \cite{boyd-nonlinear-book}. 
 The linear susceptibility $\chi_{ij}^{(1)}$ is scalar in this point group.
  According to Eq. \eqref{Eq:eta2}, $\eta_{ijk}^{(2)}$ has the same nonzero components as $\chi_{ijk}^{(2)}$. 
 Some semiconductors such as GaAs, GaP, and ZnTe are in this category, having very large $\chi^{(2)}$ compared with widely used nonlinear-optical media such as KDP and BBO.

 In this case, the $x$-polarized off-resonant pump gives the polarization entanglement as 
 \begin{align}
  &c_\mathrm{PS}|\mathrm{P}_{\vb*{k}_c}\rangle |\mathrm{S}_{-\vb*{k}_c}\rangle +  
  c_\mathrm{SP}|\mathrm{S}_{\vb*{k}_c}\rangle |\mathrm{P}_{-\vb*{k}_c}\rangle, \\ &c_\mathrm{SP}=\ee^{\ii(\lambda_{\mathrm{S}\vb*{k}_c}+\lambda_{\mathrm{P}(-\vb*{k}_c)}-\lambda_{p\vb*{0}})}c_\mathrm{PS}^*,
 \end{align}
 and the $y$-polarized off-resonant pump gives 
\begin{align} 
c_\mathrm{PP}|\mathrm{P}_{\vb*{k}_c}\rangle |\mathrm{P}_{-\vb*{k}_c}\rangle +  c_\mathrm{SS}|\mathrm{S}_{\vb*{k}_c}\rangle |\mathrm{S}_{-\vb*{k}_c}\rangle
\end{align} 
under the time-reversal symmetry, namely, 
\begin{align}
\vb*{u}_{\tilde{\alpha}\vb*{k}}^*(\vb*{x})=\ee^{\ii\lambda_{\tilde{\alpha}\vb*{k}}}
\vb*{u}_{\tilde{\alpha}(-\vb*{k})}(\vb*{x}), \quad \lambda_{\tilde{\alpha}\vb*{k}}=\lambda_{\tilde{\alpha}(-\vb*{k})}.
\end{align}

The polarization state is maximally entangled in the former case, whereas  the latter is not.  
The absolute values of complex coefficients $c_\mathrm{PP},c_\mathrm{PS}, c_\mathrm{SP},c_\mathrm{SS}$ are available classically via Eq. \eqref{Eq:flux}.
Determining their phases is essential in entanglement manipulation, but requires a detailed analysis of the eigenmodes. 
Here, we focus on the absolute values that relate to the conversion efficiency of the SPDC.

\section{Case of a monolayer of dielectric spheres}

\begin{figure}
	\includegraphics[width=0.45\textwidth]{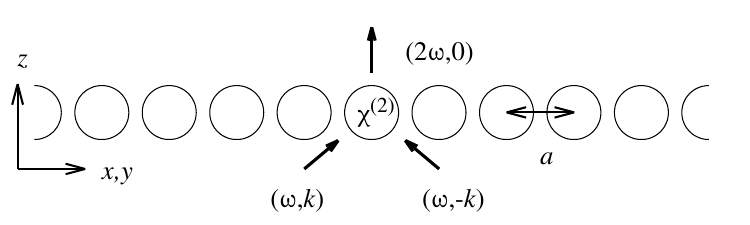}
	\caption{\label{Fig:geo} 
		Schematic illustration of the toy model system under study. Two plane-wave lights are incident on a monolayer of spheres arranged in the square lattice with lattice constant $a$, and the noncollinear SHG takes place. The spheres are made of a noncentrosymmetric material with the second-order electric susceptibility $\chi^{(2)}$.   
	}
\end{figure}

As a model calculation, let us consider the square lattice of dielectric spheres made of a noncentrosymmetric material with the $T_d$ point group. 
A schematic illustration of the system under study is shown in Fig. \ref{Fig:geo}. 
The two incident plane-wave lights with the same frequency come into the monolayer, and the noncollinear SHG is induced. The radiation flux of the SHG gives the conversion efficiency of the SPDC through Eq.  \eqref{Eq:flux}.

\begin{figure}
\includegraphics[width=0.45\textwidth]{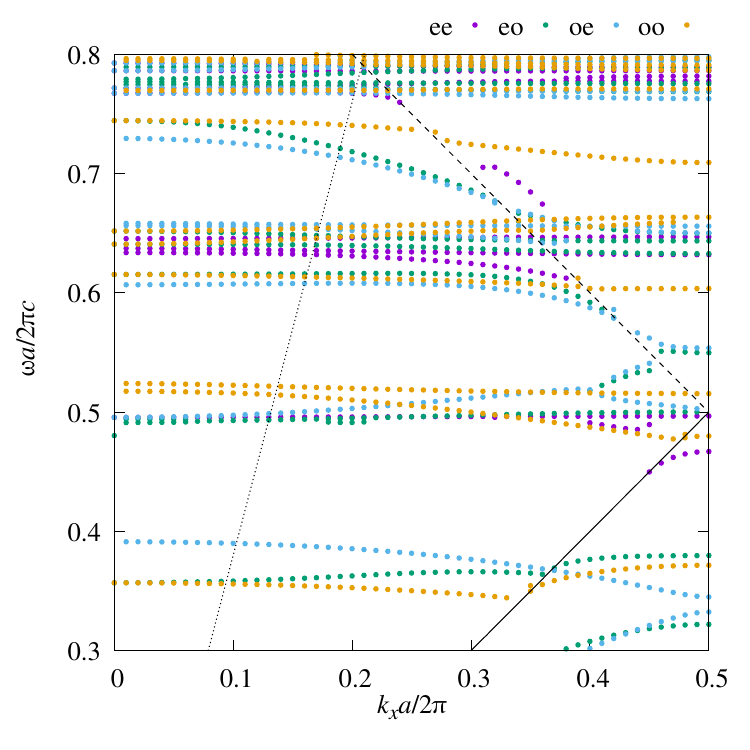}
\caption{\label{Fig:bd} 
Photonic band structure along the $\Gamma$X ($k_x$) direction in the square lattice of dielectric spheres. The spheres have dielectric constant 12 and radius $0.4a$ where $a$ is the lattice constant. The photonic band modes are classified according to the parities relevant to $y$ and $z$ directions. Symbols "ee","eo", "oe", and "oo"  stand for ($y$-even,$z$-even),  ($y$-even,$z$-odd), ($y$-odd,$z$-even), and  ($y$-odd,$z$-odd), respectively. Solid, dashed, and dotted lines represent the light cone, the diffraction threshold, and the line of constant incident angle $\theta=15.26^\circ$, respectively.  
}
\end{figure}

Figure \ref{Fig:bd} shows the photonic band structure of the true-guided and quasi-guided modes along the $\Gamma$X direction of the square lattice. 
The band structure was calculated via the photonic layer Korringa-Kohn-Rostoker (KKR) method \cite{OHTAKA::13:p667-680:1980,MODINOS::141:p575-588:1987,stefanou1992scattering}. 
Inside the light cone, the band structure is obtained by fitting the scattering phase shift $\delta$ \cite{Ohtaka:I:Y::70:p035109:2004} to the Breit-Wigner form as 
\begin{align}
\ee^{2\ii\delta}=	\ee^{2\ii\delta_0}\frac{\omega-\omega_0-\ii\gamma}{\omega-\omega_0+\ii\gamma}. 	
\end{align}
where $\delta_0$ is the background phase shift, $\omega_0$ is the resonance frequency, and $\gamma$ is its width (or inverse lifetime). Here, we plot $\omega_0$ as a function of Bloch momentum $\vb*{k}$. 
The band structure  is classified according to the parities in the 
$y$ and $z$ directions. 
It have several band crossing points between $y$-even and $y$-odd bands.  
At these crossing points $(\vb*{k}_c,\omega_c)$, the degenerate SPDC can be enhanced.

\begin{figure}
	\includegraphics[width=0.45\textwidth]{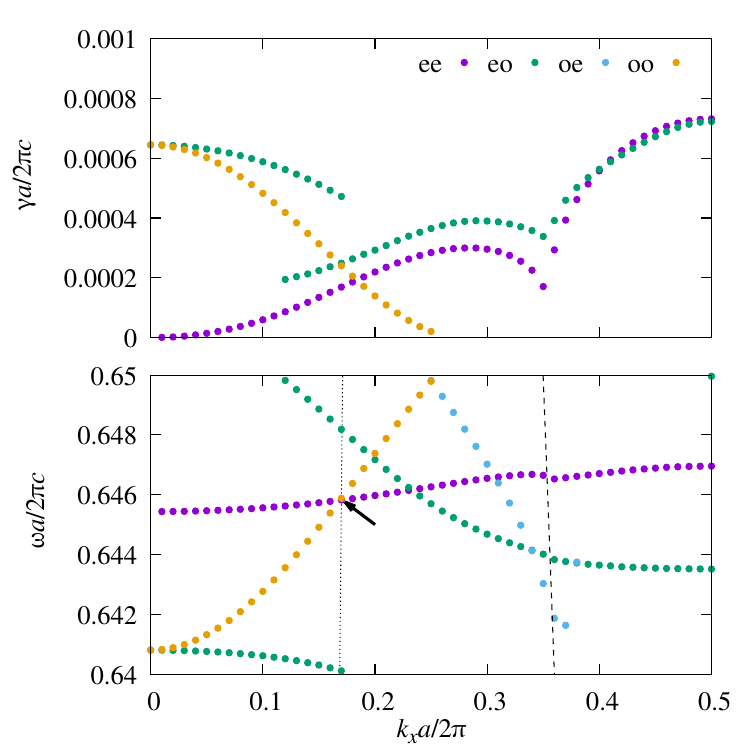}
	\caption{\label{Fig:bd2} 
		Close-up view of Fig. \ref{Fig:bd} (lower panel) together with the imaginary photonic band structure (upper panel). The arrow indicated the band-crossing point we focus on in this paper. 
	}
\end{figure}

Figure \ref{Fig:bd2} shows a close-up view of the band structure together with the resonance width $\gamma$.  
Here, we focus on the band-crossing point between "ee" and "oo" bands. They have relatively small $\gamma$, so that the light-matter interaction is enhanced
through these two modes.

Next, we consider the conversion efficiency of the SPDC via the classical calculation of the noncollinear SHG. We assume two incident plane waves with the same angular frequency $\omega/2$ and opposite incident angles $\theta_c=\pm \sin^{-1}(ck_c/\omega_c)$, which is  adjusted for a particular crossing point. The incident light  induces the noncollinear SHG of angular frequency $\omega$.  
Then, the resulting radiation field of the SHG in the far field is expressed as  
\begin{align}
&\vb*{E}^{(2)\pm}(\vb*{x})=\sum_{\vb*{g}}^\mathrm{open}
\ee^{\ii\vb*{K}_{\vb*{g}}^\pm\cdot\vb*{x}} \vb*{t}_{\vb*{g}}^{(2)\pm},\\
&\vb*{K}_{\vb*{g}}^\pm=\vb*{g}\pm\hat{z}\Gamma_{\vb*{g}},\quad 
\Gamma_{\vb*{g}}=\sqrt{\qty(\frac{\omega}{c})^2-\vb*{g}^2}.
\end{align}
where $\vb*{g}$ is a 2D reciprocal lattice vector and superscript $\pm$ refers to the sign of $z$ (the monolayer center is taken to be $z=0$).   
The plane-wave coefficient $\vb*{t}_{\vb*{g}}^{(2)\pm}$ can also be calculated via the photonic layer KKR method.  

The radiation flux of the SHG per unit area is given by 
\begin{align}
F^{(2)}=\sum_{\vb*{g}}^\mathrm{open}\frac{\Gamma_{\vb*{g}}}{2\mu_0\omega}\qty( 
 |\vb*{t}_{\vb*{g}}^{(2)+}|^2+|\vb*{t}_{\vb*{g}}^{(2)-}|^2).
\end{align}
This flux should be identified with Eq. \eqref{Eq:flux} divided by the planar area of the system.

Under a non-resonant pumping, the eigenmode relevant to the pump light is like a plane wave. Therefore, the electric field of the pump light has a little $z$ dependence inside the monolayer, provided the monolayer thickness is in the subwavelength regime. 
In this case together with the point group $T_d$, the signal and idler lights are better to have opposite parities in the $z$ direction. Otherwise, factor 
$F_{p\alpha_1\alpha_2}$ becomes small.

\begin{figure}
\includegraphics[width=0.45\textwidth]{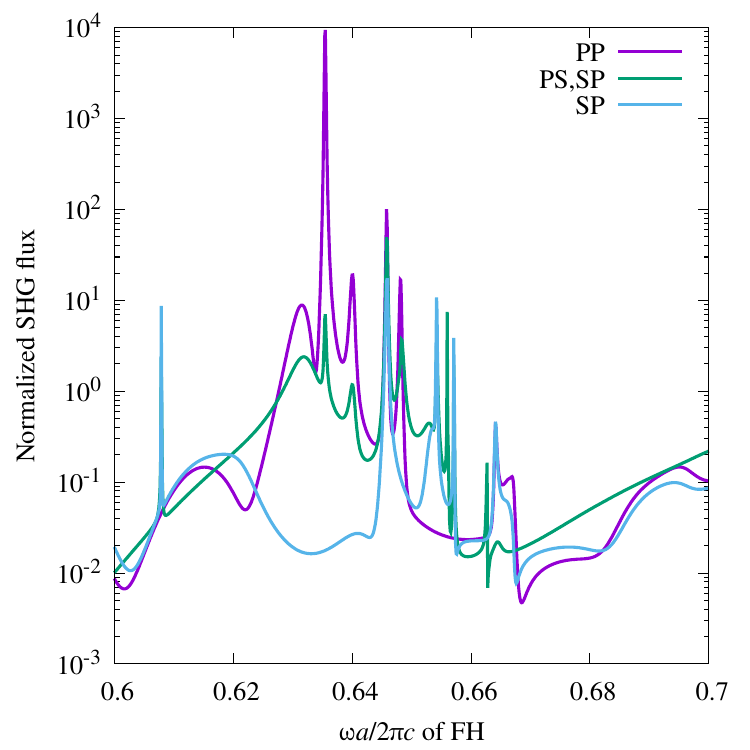}
\caption{\label{Fig:sfg} 
Normalized radiation flux of the noncollinear SHG from the monolayer of the spheres. The spheres are made of a noncentrosymmetric material with the $T_d$ point group, which has $\chi_{ijk}^{(2)}=\chi_t^{(2)}(\ne 0)$ for $(ijk)=(xyz),(xzy),(yzx),(yxz),(zxy),(zyx)$ and zero otherwise.   
The incident angle of the first-harmonic (FH) waves is fixed to be $\theta_c=\pm 15.26^\circ$
to excite the quasi-guided modes at the crossing point of Fig. \ref{Fig:bd2}. The two FH waves are either P- or S-polarized.  Symbols "PP",PS","SP", and "SS" represent the (signal,idler) lights are 
(P,P), (P,S), (S,P), and (S,S) polarized, respectively.     
Normalized radiation flux is defined by $2c\mu_0 F^{(2)}/|\chi_t^{(2)}e_0^2|^2$ 
where $e_0$ is the electric field amplitude of the FH light. 
	}
\end{figure}

Figure \ref{Fig:sfg} shows the radiation flux of the noncollinear SHG for the two incident waves of either P or S-polarization. 
The incident angle $\theta_c$ is fixed to excite the modes at a crossing point 
[$\theta_c=\pm\sin^{-1}(ck_c/\omega_c)$] and angular frequency is scanned. 
The SHG flux is strongly enhanced at various first-harmonic (FH) frequencies. 
These peaks represent the excitement of P- or S-polarized quasi-guided modes at FH frequencies. Or, the excitement of the quasi-guided modes at the $\Gamma$ point of second-harmonic frequencies. 
However, all of these peaks are not directly related to the SPDC of  interest, because the calculated flux includes those of the diffraction channels other than the non-diffractive ($\vb*{g}=0$) one of the normal direction. The normal direction is supposed to be the pumping direction of the SPDC.

\begin{figure}
	\includegraphics[width=0.45\textwidth]{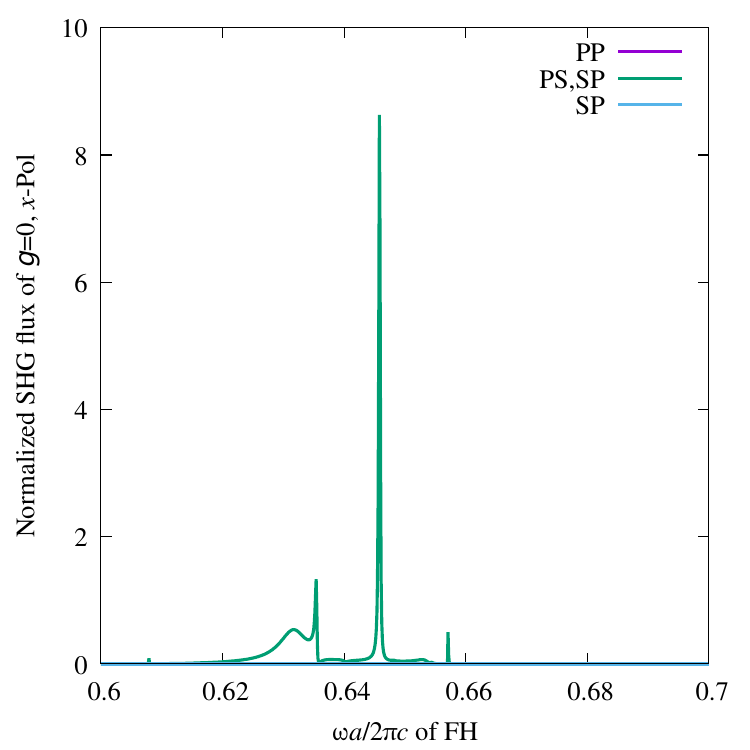}
	\caption{\label{Fig:sfgnormal} 	
	Same as in Fig. \ref{Fig:sfg}. However, the radiation flux is limited to the contribution of $(\vb*{t}_{\vb*{g}=0}^{(2)+})_x$. The fluxes of the incident PP and SS polarizations of the FH wave are very tiny (but nonzero) and are not visible in the graph scale.  
	}
\end{figure}

Instead, if we plot solely the flux contribution in the normal direction,  the result changes drastically from Fig. \ref{Fig:sfg}, as shown in Fig. \ref{Fig:sfgnormal}. 
There is a remarkable peak only for (signal,idler)=(P,S) and (S,P) at $\omega a/2\pi c=0.646$, where the P- and S-polarized bands cross as in Fig. \ref{Fig:bd2} and the SHG light is dominantly $x$-polarized as expected. 
The $y$-polarized SHG component is negligible in the entire frequency range of Fig. \ref{Fig:sfgnormal}.  
However, if we closely look at the small peak of the crossing point concerned, we can observe that the $y$-polarized SHG component is enhanced only for  (signal,idler)=(P,P) or (S,S), as expected from the selection rule in Sec. IV.   
In this case, most power is transmitted to the four diffraction channels with $\vb*{g}=\pm\hat{x},\pm \hat{y}$ $(\times 2\pi/a)$. Therefore, the polarization entanglement is limited only if we send the pump light from the four different angles that correspond to the above diffraction channels.  
It is better to design the resonance such that $2\omega_c$ is below the diffraction threshold $2\pi c/a$.

\begin{figure}
	\includegraphics[width=0.45\textwidth]{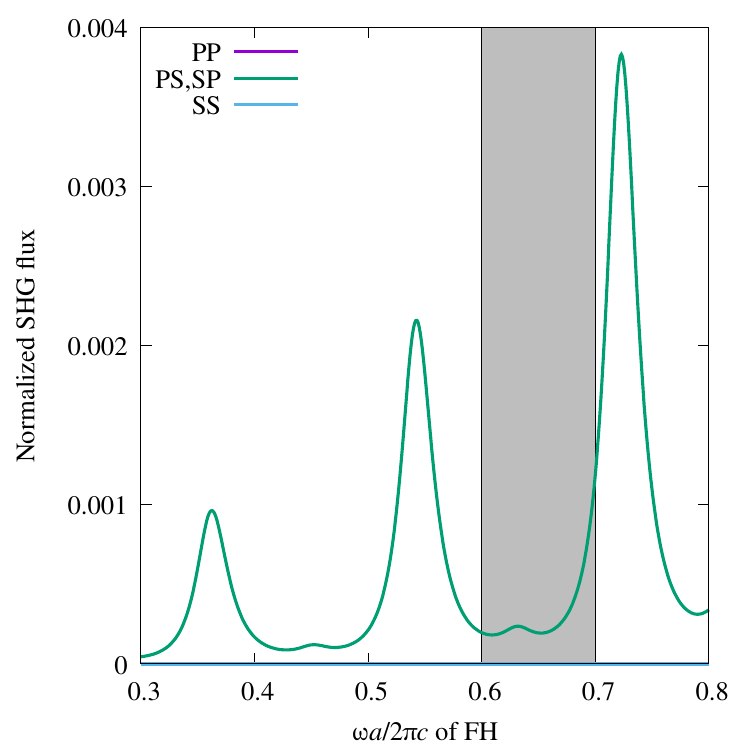}
	\caption{\label{Fig:sfg_slab} 
		Normalized radiation flux of the noncollinear SHG from the uniform slab with the same material as in Fig. \ref{Fig:sfg}.   The setting of the incident wave is also the same. The slab thickness is taken to be $0.8a$. 
		The gray region corresponds to the frequency interval of Figs. \ref{Fig:sfg} and \ref{Fig:sfgnormal}. The SHG flux for (signal,idler)=(P,P) and (S,S) vanishes. 
	}
\end{figure}

For comparison, we show in Fig. \ref{Fig:sfg_slab} the radiation flux of the noncollinear SHG from a uniform slab made of the same material as in Fig. \ref{Fig:sfg}. The thickness of the slab is taken to be the same as the diameter of the spheres. 
There is no marked signal other than the three broad peaks for (signal,idler)=(P,S) and (S,P).  They correspond to the Fabri-Perot resonance of the slab. Moreover, the normalized SHG flux is much smaller than in Figs. \ref{Fig:sfg} and \ref{Fig:sfgnormal}. 
As for (signal,idler)=(P,P) and (S,S), the SHG flux is identically zero by symmetry.

\section{Summary and discussion}
In summary, we have presented a theoretical analysis of the degenerate SPDC in nonlinear metasurfaces. A band crossing between P- and S-polarized quasi-guided modes on a mirror axis yields a boosted SPDC with a polarization-entangled biphoton state.  The conversion efficiency in the SPDC is available classically via the inverse process of the noncollinear SHG. We demonstrate these features 
in a monolayer of noncentrosymmetric spheres arranged in the square lattice. 

There are many issues that remain to be investigated. One is to determine and design the relative phase between $|\mathrm{P}\rangle|\mathrm{S}\rangle$ and $|\mathrm{S}\rangle|\mathrm{P}\rangle$ (or $|\mathrm{P}\rangle|\mathrm{P}\rangle$ and $|\mathrm{S}\rangle|\mathrm{S}\rangle$) mentioned in the text. It requires a detailed analysis of the relevant eigenmodes and time-reversal symmetry.

Another critical issue is an on-resonant pumping. To make the system simple enough, we have assumed the off-resonant pumping. There, the pump light is plane-wave like even inside the metasurfaces. 
The on-resonant pump implies the excitation of a quasi-guided mode at the $\Gamma$ point. The local field of the quasi-guided mode is no longer plane-wave like, so that the selection rule becomes complicated, mixing all the four combinations of  $|\mathrm{P}\rangle|\mathrm{P}\rangle$,  $|\mathrm{P}\rangle|\mathrm{S}\rangle$, $|\mathrm{S}\rangle|\mathrm{P}\rangle$, and  $|\mathrm{S}\rangle|\mathrm{S}\rangle$. 
At the same time, the on-resonant pump further boosts the conversion efficiency 
of the SPDC.  

It is also essential to implement the present scheme of the SPDC  in   experimentally more accessible platforms, e.g., dielectric slabs with a periodic array of air holes.   Since the present theory assumes a mirror symmetry and the time-reversal symmetry as a minimal requirement, there is no obstacle to applying the scheme. In the theory side, we need to extend, for instance, the rigorous coupled wave analysis \cite{moharam1981rigorous}, to deal with the noncollinear SHG.

We hope this paper stimulates further investigation of the SPDC in nonlinear metasurfaces.


\begin{acknowledgments}

This work was supported by JSPS KAKENHI Grant No. 22K03488. 
	
\end{acknowledgments}

%

\end{document}